\documentclass[11pt,a4paper]{article}

\usepackage{cite}

\usepackage[utf8x]{inputenc}
\usepackage[english]{babel}
\usepackage{hyperref}
\usepackage{lipsum}
\usepackage{amsmath, amssymb,slashed,mathrsfs, bbm, braket, color}

\newcommand{\SO}{\text{SO}}

\numberwithin{equation}{section}

\newcommand{\eg}{{\emph{e.g. }}}
\newcommand{\id}{{\mathbbm 1}}
\newcommand{\im}{{\rm i}}
\newcommand{\ads}{{\rm AdS}}
\newcommand{\p}[1]{\left(#1\right)}
\newcommand{\comm}[1]{\left[#1\right]}
\newcommand{\acomm}[1]{\left\{#1\right\}}
\newcommand{\gcomm}[1]{\left[#1\right\}}
\newcommand{\f}[2]{\frac{#1}{#2}}
\newcommand{ \bb}[1]{{\mathbb #1}}
\newcommand{ \gen}[1]{\mathfrak{#1}}

\def\be{\begin{equation}}
\def\ee{\end{equation}}

\def\ba{\begin{align}}
\def\ea{\end{align}}

\def\A{\mathcal A}
\def\Y{\mathcal Y}

\begin{document}
\begin{titlepage}
\begin{flushright}%\footnotesize\ttfamily
UUITP -- 40/17
\end{flushright}
\vskip 1.5in
\begin{center}
{\bf\Large{Yangian Symmetry of String Theory on ${\rm AdS}_3\times S^3\times S^3\times S^1 $ with Mixed 3-form Flux}}
\vskip
0.5in {Antonio Pittelli } \vskip 0.5in {\small{ \emph{Department of Physics and Astronomy, Uppsala University,\\ Box 516, SE-75120 Uppsala, Sweden} }
}
\end{center}
\vskip 0.5in
\baselineskip 16pt
\begin{abstract}  

We find the Yangian  symmetry underlying the integrability of type IIB superstrings on $\ads_3\times S^3\times S^3\times S^1 $ with mixed Ramond-Ramond  and Neveu-Schwarz-Neveu-Schwarz  flux.  The abstract commutation relations of the Yangian are formulated via RTT realisation, while its matrix realisation is in an evaluation representation depending  on the quantised coefficient of the Wess-Zumino term.  The construction naturally encodes a secret symmetry of the worldsheet scattering matrix whose generators map different Yangian levels to each other. We  show that in the large effective string tension limit  the Yangian becomes a deformation of a unitary loop algebra and we derive its universal  classical r-matrix.

 \end{abstract}
 
% \bigskip
%\begin{center}
%{\bf Keywords}: Quantum algebra, integrable systems, AdS/CFT
%\end{center}

\date{}
\end{titlepage}

\tableofcontents

\section{Introduction}

\subsection{Yangian Symmetry and Exact Solvability }

There are more symmetries in a theory than those displayed by its Lagrangian. This is particularly evident in four dimensional $\mathcal N=4$ Super Yang-Mills theory, whose planar on-shell scattering amplitudes \cite{Drummond:2009fd} and Wilson loops \cite{Muller:2013rta} are invariant under an infinite tower of non-local conserved charges. The very same symmetry, identified with    the \emph{Yangian} $\Y[\gen{psu}(2,2|4)]$, was found in type IIB superstrings on $\ads_5\times S^5$ by studying  the coset structure of the background \cite{Bena:2003wd}. Generally, if $\gen g$ is a graded Lie algebra, the corresponding  Yangian $\Y[\gen g]$  is a particular quantum deformation of $\mathcal U(\gen g[u])$,  the universal enveloping algebra of   $\gen g$-valued polynomials in the  spectral parameter $u\in\bb C$.  In terms of \emph{Drinfeld first realisation}  \cite{Takhtajan:1979iv,Drinfeld:1985rx,Drinfeld:1987sy}, $\Y[\gen g]$ is formulated as  
\be\label{eq: dfr}
\Y[\gen g]=\bigcup_{n\in\bb N} \Y_n[\gen g], \quad \Y_n[\gen g]={\rm span }  \;\gen J^A_{(n)},\quad \gcomm{\gen J^A_{(m)},\gen J^A_{(n)}}={f^{AB}}_C\, \gen J^C_{(m+n)},\quad m+n=0,1;
\ee
where ${f^{AB}}_C$ are the structure constant of $\gen g$, $[X,Y\}=XY-(-)^{|X||Y|}YX$ is the graded commutator of $X$ and $Y$ and $|\cdot|$   the Gra\ss mann grading. In particular, the Lie algebra $\gen g$ coincides with $\Y_0[\gen g]$. Moreover, $\Y[\gen g]$   carries  a Hopf algebra structure and its generators satisfy   generalised  Jacobi identities called \emph{Serre relations}. All   $\Y_n[\gen g]$ with $n>1$ can be reconstructed  out of   lower levels via (\ref{eq: dfr}): the whole $\Y[\gen g]$ is then determined by $\Y_0[\gen g]$ and $\Y_1[\gen g]$. Eventually,   $\Y[\gen g]$ admits an \emph{evaluation representation} if there exists a spectral parameter dependent map $\rho_u$ such that
\be
\rho_u\, \gen J^A_{(n)}=u^n\, \gen J^A_{(0)},\qquad J^A_{(0)}\in\gen g, \qquad n\in\bb N.
\ee

Yangian symmetry  typically appears in integrable quantum field theories: indeed, both  $\mathcal N=4$ super Yang-Mills theory \cite{Minahan:2002ve,Beisert:2005fw, Beisert:2005tm} and type IIB superstrings on $\ads_5\times S^5$ \cite{Arutyunov:2003uj,Arutyunov:2006ak,Arutyunov:2006yd} are dual to an integrable spin chain system. This duality is even more powerful than the AdS/CFT correspondence \cite{Maldacena:1997re, Gubser:1998bc, Witten:1998qj} as the integrability picture allows for computing the conformal data of $\mathcal N=4$ super Yang-Mills theory at all orders in the gauge coupling constant  \cite{Basso:2015zoa, Eden:2015ija,Basso:2015eqa, Eden:2017ozn}. Technically, it is therefore very useful to have the control on the dual integrable picture  through its symmetries, as the latter are able to strictly constrain or even determine gauge theory or gravity observables. Conceptually, Yangians are fascinating because they link to each other  completely different models such as gauge, gravity and   condensed matter systems via a rich  mathematical structure. Yangians have been found in quite a few instances of the AdS/CFT correspondence \cite{Beisert:2014hya, Pittelli:2014ria, Hoare:2015kla, Borsato:2016xns}. In this paper we will focus on type IIB superstrings on $\ads_3\times S^3\times S^3\times S^1 $ with mixed Ramond-Ramond (RR)  and Neveu-Schwarz-Neveu-Schwarz (NSNS)  flux. We shall now review the main features of the model.

%%%%%%%%%%%%%%%%%%%%%%%

\subsection{Superstrings on $\ads_3\times S^3\times S^3\times S^1$ with Mixed 3-Form Flux}

The near-horizon geometry of two M5 branes  intersecting an M2 brane is $\ads_3\times S^3\times S^3\times T^2$, whose dimensional reduction along $S^1$ gives type IIA superstrings on $\ads_3\times S^3\times S^3\times S^1$ with 16 supersymmetries. By performing T-duality along the $S^1$ one obtains the corresponding type IIB model \cite{Grisaru:1985fv, Cowdall:1998bu,Boonstra:1998yu,Gauntlett:1998kc,Elitzur:1998mm,deBoer:1999gea}. $\ads_3\times S^3\times S^3\times S^1$ supports both RR and NSNS fluxes and displays ${ \rm D}(2,1;\alpha)\times {\rm  D}(2,1;\alpha)\times {\rm U}(1)$ isometry, with the AdS and sphere radii  being   parametrised by $\alpha$:
\be
\alpha=R^2_\ads/R^2_{S_+^3},\qquad 1-\alpha=R^2_\ads/R^2_{S_-^3}, 
\ee
where the spheres are labelled by $\pm$. If the spheres have equal radii, $\alpha=1/2$ and  the exceptional superalgebra ${\gen d}(2,1;\alpha)$ becomes $\gen{osp}(4|2)$. Instead, if the AdS radius is equal to that of one of the spheres, the other sphere can be compactified into a $T^3$ giving the $\ads_3\times S^3\times T^4$ background, whose complete worldsheet S-matrix was found in \cite{Borsato:2014hja}. In this limit, $\alpha=0$ or $\alpha=1$ and   ${\gen d}(2,1;\alpha)$ reduces to $\gen{psu}(2|2)$. The massive sector of the worldsheet scattering matrix of type IIB strings on $\ads_3\times S^3\times S^3\times S^1$ with RR flux was first  derived in \cite{Borsato:2012ud}, while its integrability properties were studied in \cite{Babichenko:2009dk} exploiting the coset structure of the background\footnote{Such a coset structure was also used to prove the T-self duality of $\ads_3\times S^3\times S^3\times S^1$ in \cite{Abbott:2015mla}. T-self duality is related to dual superconformal symmetry, which is a subsector of the full Yangian algebra \cite{Berkovits:2008ic, Beisert:2008iq}.}, 
\be
\ads_3\times S^3\times S^3\times S^1=\f{{\rm D}(2,1;\alpha)\times {\rm D}(2,1;\alpha)}{\SO(1,2)\times \SO(3)\times \SO(3)}.
\ee
The investigation was generalised by including  both RR and NSNS fluxes in \cite{Borsato:2015mma} directly working on the Green-Schwarz action of the theory,  ignoring the  coset formulation. In light-cone gauge, the worldsheet scattering is encoded  in an $\gen{su}_c(1|1)^2$ invariant R-matrix, which was derived assuming the integrability of the quantum theory. As in the $\ads_5\times S^5$ case, to such an integrability shall correspond a Yangian symmetry restricting the observables of the     conformal field theory dual, first tackled in \cite{Gukov:2004ym,Tong:2014yna} and deeply studied in \cite{Baggio:2017kza, Eberhardt:2017fsi, Eberhardt:2017pty }. The goal of this paper is indeed to extract the Yangian of $\ads_3\times S^3\times S^3\times S^1$ superstrings with mixed flux. Especially, we will employ the scattering matrix found in \cite{Borsato:2015mma} to define its own symmetry algebra. The procedure we will use is called RTT realisation \cite{Faddeev:1987ih,Kulish:1980ii,Molev:1994rs}: we introduce it in the next subsection.

%%%%%%%%%%%%%%%%%%%%%%%

\subsection{Integrable Scattering matrices from Hopf Algebras}

Integrable quantum field theories enjoy an infinite number of conserved charges. Such a   symmetry, which we denote by $\mathcal A$, is so constraining that it completely fixes the scattering matrix of the system. This situation is rigorously  described if $\mathcal A$ has got a Hopf algebra structure. Indeed, the coproduct map $\Delta:\mathcal A\to \mathcal A\otimes \mathcal A$ naturally provides a multiparticle representation of the conserved charges: conventionally,  $\gen J\in\mathcal A$ acts on in-states via $\Delta(\gen J)$ and on out-states via $\Delta^{\rm op}(\gen J)=\mathcal P\circ \Delta(\gen J)$, where $\mathcal P(X\otimes Y)=(-1)^{|X||Y|}Y\otimes X$ is the graded permutation operator. The symmetry acts on antiparticle states  through the antipode $\Sigma:\mathcal A	\to \mathcal A$, which is a $\bb C$-linear anti-homomorphism.

This provided, the scattering matrix $S$ acting on the Hilbert space $\mathscr H$ can be expressed as $S=\mathcal P\circ R$, with  $  R:\bb C\times\bb C\to {\rm End}(\mathscr H\otimes \mathscr H)$  intertwining  $\Delta$ and  $\Delta^{\rm op}$. The R-matrix depends on two spectral parameters $u_1,u_2\in\bb C$ related to the momenta of the scattering excitations and  is fully determined by  the \emph{quasi co-commutativity condition}
\begin{equation}\label{eq: QuasiCocommutativity}
\Delta^{\rm op}(\gen J)\,R=R\,\Delta(\gen J),\qquad \forall\,\gen J\in \mathcal A,
\end{equation} 
 as well as by the \emph{crossing equations}
\begin{equation}\label{eq: CrossingEquations}
(\Sigma\otimes\id)\,R=(\id\otimes\Sigma )\,R=R^{-1}.
\end{equation}
Specifically, (\ref{eq: CrossingEquations}) fixes the \emph{dressing phases}  left unconstrained by  (\ref{eq: QuasiCocommutativity}), once the analyticity properties of such phase factors are specified\footnote{The dressing phase for $\ads_5/{\rm CFT}_4$ was studied in \cite{Arutyunov:2004vx, Janik:2006dc,Beisert:2006ib,Beisert:2006ez}, while the  $\ads_3/{\rm CFT}_2$ case was investigated in \cite{Borsato:2013hoa, Borsato:2016xns}.}. Then, the S-matrix immediately follows from $S=\mathcal P\circ R$. Importantly,  (\ref{eq: QuasiCocommutativity}) implies  that, if $\gen C\in\mathcal A$ is central, it must be co-commutative: $\Delta^{\rm op}(\gen C)=\Delta (\gen C)$. We shall also require  that the underlying Hopf algebra $\mathcal A$ is \emph{almost quasi triangular}, meaning   that the \emph{fusion relations}\footnote{We denote by   $t_{ij}$ a tensor acting upon the $i$-th and $j$-th components of the multiple tensor product of representations $V_1\otimes\cdots\otimes V_i\otimes\cdots\otimes V_j \otimes\cdots $. }
\begin{equation}\label{eq: FusionRelations}
\p{\Delta\otimes\id}R_{12}=R_{13}\,R_{23},\qquad \p{\id\otimes\Delta }R_{12}=R_{13}\,R_{12}
\end{equation}
hold. This is not restrictive as (\ref{eq: FusionRelations}) and (\ref{eq: QuasiCocommutativity}) imply the \emph{quantum Yang-Baxter equation} (QYBE) \cite{Baxter:1972hz}
 \begin{equation}\label{eq: qybe}
R_{12}\,R_{13}\,R_{23}=R_{23}\,R_{13}\,R_{12},
\end{equation}
which is a necessary and characterising  condition for  R-matrices  of   integrable systems.

%%%%%%%%%%%%%%%%%%%%%%%

\subsection{From R-matrices to Hopf Algebras}

The above discussion   can be read backwards; indeed,  an R-matrix satisfying the QYBE defines its own Hopf algebra structure by  the \emph{RTT relations} 
\begin{equation}\label{eq: RTTrelations}
R_{12}(u_1,u_2)\,\mathcal T_1(u_1)\,\mathcal T_2(u_2)=\mathcal T_2(u_2)\,\mathcal T_1(u_1)\,R_{12}(u_1,u_2),
\end{equation}
where  the monodromy matrix $\mathcal T:\bb C\to {\rm End}(\mathscr H)\otimes\mathcal A$  generates the conserved charges. If we focus on the  $\ads_3\times S^3\times S^3\times S^1$ worldsheet scattering,   ${\rm End}(\mathscr H)=\gen{su}_c(1|1)^2\subset\gen{gl}(2|2)$ and we can    express the R-matrix and the monodromy matrix in the standard basis\footnote{Such a basis comprehends 16 matrices with components  ${\p{{e^a}_b}^i}_j=(-1)^{|b|}\delta^{ia}\,\delta_{bj}$ and $a,b,i,j=1,\dots,4$. We choose the grading of the indices to be $|1|=|2|=-|3|=-|4|=+1$, meaning that 1,2 are   bosonic indices while 3,4 are fermionic ones.
}  $\acomm{{e^a}_b}$ of the $\gen{gl}(2|2)$ superalgebra. This  embedding will be particularly convenient as it allows for dealing with both copies of $\gen{su}_c(1|1)$ at the same time. Then, assuming  that  $\mathcal T(u)$ is holomorphic in a neighbourhood of $u=\infty$, 
\begin{equation}\label{eq: Tmatrixglbasis}
\mathcal T(u)=\sum_{n\in\bb N}\sum_{a,b=1}^4u^{-n}(-1)^{|b|}{e^b}_a\otimes \mathbb T^{\,a}_{(n-1)\,b},
\end{equation}
with $\mathbb T^{\,a}_{(n)\,b}$ being the abstract (representation independent) generators of $\mathcal A$. Similarly, the   R-matrix takes the form 
\begin{equation}\label{eq: Rmatrixglbasis}
 R =\sum_{a,b,c,d=1}^4 (-1)^{|b|+|c|}R^{bd}_{ac}(u_1,u_2)\,{e^a}_b\otimes{e^c}_d.
\end{equation}
Substituting  (\ref{eq: Tmatrixglbasis}) and (\ref{eq: Rmatrixglbasis}) in  (\ref{eq: RTTrelations})   yields
\begin{align}\label{eq: RTTRelationsComponents}
&\sum_{n_1\in\bb N}\sum_{n_2\in\bb N} \sum_{e=1}^4\sum_{f=1}^4 \,u_1^{-n_1}u_2^{-n_2}\left[(-1)^{(|c|+|e|)(|b|+|d|)+|f|}\, R^{ef}_{ab}\,\bb  T^{\,c}_{(n_1-1)\,e}\bb  T^{\,d}_{(n_2-1)\,f}\right.+\nonumber\\
&\left.-(-1)^{(|b|+|d|)(|d|+|f|)+|b|} \, R^{cd}_{ef}\,\bb  T^{\,f}_{(n_2-1)\,b}\bb  T^{\,e}_{(n_1-1)\,a}\right]=0.
\end{align}
The abstract commutation relations of $\mathcal A$ are therefore recovered by expanding this constraint around $u_1,u_2=\infty$: for instance,  (\ref{eq: RTTRelationsComponents})  implies    $\mathbb T^{\,a}_{(-1)\,b}=\bb U\,{\delta^a}_b$, where $\bb U$ is a central element. Furthermore, by comparing the RTT relations (\ref{eq: RTTrelations}) with the QYBE (\ref{eq: qybe}), one observes   that $R$ and  $\mathcal T(u)$ can be connected via  a representation map $\rho_u$ depending on the  spectral parameter $u$:    $R(u_1,u_2)=\p{\id\otimes\rho_{u_2}}\mathcal T(u_1)$. As a result, the formula
\begin{align}\label{eq: RTTRepsfromRmatrix}
 \sum_{c=1}^4\sum_{d=1}^4(-1)^{|c|}R^{ac}_{bd}(u_1,u_2)\,{e^d}_c =T^{\,a}_{(-1)\,b}+\f1{u_1}\,T^{\,a}_{(0)\,b} +\f1{u_1^2}\,T^{\,a}_{(+1)\,b} +\dots
\end{align} 
gives access to the representations   $T^{\,a}_{(n)\,b}=\rho_{u}\bb T^{\,a}_{(n)\,b}$. As for the Hopf algebra structure, the fusion relations (\ref{eq: FusionRelations}) induce the coproducts 
\begin{align}
\Delta\p{\bb U}&=\bb U\otimes \bb U,\qquad \Delta\p{\bb T^{\,a}_{(0)\,b}}=\bb T^{\,a}_{(0)\,b}\otimes \bb U^{|b|}+\bb U^{|a|}\otimes \bb T^{\,a}_{(0)\,b},\nonumber\\
\Delta\p{\bb T^{\,a}_{(1)\,b}}&=\bb T^{\,a}_{(1)\,b}\otimes \bb U^{|b|}+\bb U^{|a|}\otimes \bb T^{\,a}_{(1)\,b}+\sum_{c=1}^4\bb T^{\,a}_{(0)\,c}\otimes \bb T^{\,c}_{(0)\,b},
\end{align}
while the antipodes descend from (\ref{eq: CrossingEquations}):
\begin{align}
 \Sigma\comm{\bb T^{\,a}_{(-1)\,b}}&=\bb U^{-|b|}{\delta^a}_b,\qquad \Sigma\comm{\bb T^{\,a}_{(0)\,b}}=-\bb U^{-|a|-|b|}\bb T^{\,a}_{(0)\,b},\nonumber\\
\Sigma\comm{\bb T^{\,a}_{(1)\,b}}&=-\bb U^{-|a|-|b|}\bb T^{\,a}_{(1)\,b}+\sum_{c=1}^4\bb U^{-|a|-|b|-|c|}\bb T^{\,a}_{(0)\,c}\bb T^{\,c}_{(0)\,b}.
\end{align}
If the Hopf algebra of the system is a Yangian over a graded Lie algebra $\gen g$, $\mathcal Y[\gen g]$, it is useful to define the objects 
\be 
 \bb J^{\,a}_{(0)\,b}:=-\f{\im h}{2  m}\, \bb U^{-|b|}  \, \bb T^{\,a}_{(0)\,b}, \qquad \bb J^{\,a}_{(1)\,b}:=-{ \f{\im h}{2 m}\, \bb U^{-|b|}  \bb T^{\,a}_{(1)\,b}-\f{\im m}h\sum_{c=1}^4(-1)^{(|a|+|c|)(|b|+|c|)} \bb J^{\,c}_{(0)\,b}  \,\bb J^{\,a}_{(0)\,c}},
\ee
where the constants  $h,m$ appear for future convenience. These $\bb Js$ satisfies 
\begin{align}\label{eq: Jcoproduct}
\Delta\p{\bb J^{\,a}_{(0)\,b}}&=\bb J^{\,a}_{(0)\,b}\otimes \id+\bb U^{|a|-|b|}\otimes \bb J^{\,a}_{(0)\,b},\nonumber\\
\Delta\p{\bb J^{\,a}_{(1)\,b}}&=\bb J^{\,a}_{(1)\,b}\otimes \id+\bb U^{|a|-|b|}\otimes \bb J^{\,a}_{(1)\,b}+\f{\im m}h\sum_{c=1}^4\bb U^{|c|-|b|}\,\bb J^{\,a}_{(0)\,c}\otimes \bb J^{\,c}_{(0)\,b}+\nonumber\\
&-\f{\im m}h\sum_{c=1}^4(-1)^{(|a|+|c|)(|c|+|b|)}\bb U^{|a|-|c|}\,\bb J^{\,c}_{(0)\,b}\otimes \bb J^{\,a}_{(0)\,c},
\end{align}
 which are the coproducts of $\mathcal Y[\gen g]$ in Drinfeld first realisation. Moreover, 
\begin{align}\label{eq: Jantipode}
\Sigma\comm{\bb J^{\,a}_{(0)\,b}}=-\bb U^{|b|-|a|}\bb J^{\,a}_{(0)\,b},\qquad \Sigma\comm{\bb J^{\,a}_{(1)\,b}}=-\bb U^{|b|-|a|}\p{\bb J^{\,a}_{(1)\,b}-\f{\im m}h\sum_{c=1}^4 \left[\bb J^{\,a}_{(0)\,c},\bb J^{\,c}_{(0)\,b}\right\}}.
\end{align}
A great advantage of the RTT realisation is that it automatically provides the         consistency conditions of the Hopf algebra, for instance  \cite{Chari:1994pz, Kassel:1995xr}
\be
\mu\circ\p{\Sigma\otimes\id}\circ\Delta\p{\gen J}=\mu\circ\p{\id\otimes\Sigma  }\circ\Delta\p{\gen J}=\epsilon\p{\gen J}\id,\qquad \gen J\in\A,
\ee 
where $\mu$ and $\epsilon$ are the Hopf algebra multiplication and counit respectively.

%%%%%%%%%%%%%%%%%%%%%%%

\subsection{Outline}

To make the paper self-contained, in Section \ref{sec: AdS3Rev} we re-derive   the   $\ads_3\times S^3\times S^3\times S^1$ superstring worldsheet scattering matrix, originally found in  \cite{Borsato:2015mma}. In  Section \ref{sec: RTTReal} we use the RTT realisation to extract  its  Yangian symmetry. We also derive the Yangian evaluation representation and show its dependence on the quantised coefficient of the Wess-Zumino term appearing in the $\ads_3\times S^3\times S^3\times S^1$ superstring action. In   Section \ref{sec: Classicalrmatrix} we perform the  large effective string tension  limit and demonstrate that the resulting classical r-matrix can be written in a universal, representation independent form as a tensor product of $\gen{u}(1|1)[u,u^{-1}]^2$ loop algebra generators.

%%%%%%%%%%%%%%%%%%%%%%%

\subsection{Note }
While completing this paper I became aware of \cite{Borsato:2017icj}, which has some overlap with this work. I am very grateful to the authors for sharing their draft before   publication.

%%%%%%%%%%%%%%%%%%%%%%%
%%%%%%%%%%%%%%%%%%%%%%%

\section{$\ads_3\times S^3\times S^3\times S^1$ R-matrix from Hopf Algebra}\label{sec: AdS3Rev}

\subsection{Centrally Extended $\gen{su}_c(1|1)^2$ Algebra and its Representations}

As we already mentioned, the fundamental symmetry of the worldsheet theory in light-cone gauge is $\gen{su}_c(1|1)^2$, which is made of two copies of $\gen{su}(1|1)$ and two central charges entangling them,
\begin{equation}
\gen{su}_c(1|1)^2=\comm{\gen{su}(1|1)_L\oplus \gen{su}(1|1)_R}\ltimes \bb R^2,
\end{equation} 
 where we labelled each copy by left (L) and right (R). Each $\gen{su}(1|1)_A$ is a three-dimensional superalgebra generated by two supercharges $Q_A,S_A$ and a central  charge   $H_A=\acomm{Q_A,S_A}$. Consequently, the algebra $\gen{su}_c(1|1)^2$ has got 4 fermionic generators $Q_L,S_L,Q_R,S_R$ and 4 central charges $H_L,H_R,P,K$ whose commutation rules are 
\begin{equation}\label{eq: AlgebraCommutationRelations}
\acomm{Q_A,S_B}=\delta_{AB}\,H_B,\qquad \acomm{Q_L,Q_R}=P,\qquad \acomm{S_L,S_R}=K;\qquad A,B=L,R.
\end{equation}

The elementary excitations of the   $\ads_3\times S^3\times S^3\times S^1$    worldsheet scattering are represented by  two bosonic states $\ket {\phi_A}$ and two fermionic ones $\ket {\psi_A}$ with $A=L,R$. The action of $Q_A,S_A$   on the   left module $\nu_L=\acomm{\ket{\phi_L},\ket{\psi_L}}$ is
\begin{align}\label{eq: LeftSingleParticleRep}
Q_L\ket {\phi_L}=a_L\ket {\psi_L},\qquad S_L\ket {\psi_L}=c_L\ket {\phi_L}, \qquad S_R\ket {\phi_L}=d_L\ket {\psi_L},\qquad Q_R\ket {\psi_L}=b_L\ket {\phi_L},
\end{align}
while the value of the central charges on $\nu_L$ is
\begin{equation}\label{eq: LeftSingleParticleCentralRep}
H_L=a_Lc_L, \qquad  H_R=b_Ld_L, \qquad P=a_Lb_L, \qquad K=c_Ld_L.
\end{equation}
The action on the right module $\nu_R=\acomm{\ket{\phi_R},\ket{\psi_L}}$ follows from LR symmetry; namely, it is obtained by substituting L with R in (\ref{eq: LeftSingleParticleRep}) and (\ref{eq: LeftSingleParticleCentralRep}): for instance, $Q_R\ket {\phi_R}=a_R\ket {\psi_R}$ and $ S_L\ket {\phi_R}=d_R\ket {\psi_R}$.  Such a representation satisfies   the shortening condition $(H_L+H_R)^2= (H_L-H_R)^2+ 4\,PK $ and is  conveniently written in terms of Zhukovski variables $\acomm{x_{L}^\pm,x_{R}^\pm}$. These are kinematical variables depending on the momentum $p$ and the mass $m$ of the corresponding scattering excitation, as well as on the effective string tension $h$ and the quantised coefficient $\kappa$ of the Wess-Zumino term in the $\ads_3\times S^3\times S^3\times S^1$ superstring action. Specifically, $\acomm{x_{L}^\pm,x_{R}^\pm}$ are defined by the constraint 
\begin{equation}\label{eq: ZhukovskyConstr}
x^+_{A}+\f1{x^+_{A}}-x^-_{A}-\f1{x^-_{A}}-\f{2\,\kappa_A}{h}\log\p{\f{x^+_{A}}{x^-_{A}}}=\f{2\im m}h,\qquad A=L,R;
\end{equation}
 where  $\kappa_L=-\kappa_R = |\kappa|$ and we chose the branch cut such that $\log e^{2\pi\im }=0$.  The link with the particle momentum is given by ${{x^+_{L}}/{x^-_{L}}}=  {{x^+_{R}}/{x^-_{R}}}=e^{\im p}$. To simplify the R-matrix entries, we define
\begin{equation}
U:=\sqrt{{x^+_{L}}/{x^-_{L}}}=\sqrt{{x^+_{R}}/{x^-_{R}}}=e^{\im p/2} ,\qquad \eta_{A}:=\p{\f{x^+_{A}}{x^-_{A}}}^{1/4}\sqrt{\f{\im h}2(x^-_{A}-x^+_{A})},
\end{equation}
whose inverse relations read
\begin{equation}
x^+_{A}=\f{2\im \,U\, \eta_{A}^2}{h\p{U^2-1}},\qquad x^-_{A}=\f{2\im \, \eta_{A}^2}{h\,U\p{U^2-1}},\qquad A=L,R.
\end{equation}
In this setting, the representation coefficients assume the compact form
\begin{align}
a_A&=\eta_A,\qquad b_A=-\f{\eta_A}{U x^-_A},\qquad c_A=\f{\eta_A}{U},\qquad d_A=-\f{\eta_A}{x^+_A};\qquad A=L,R.
\end{align}

The Lie algebra $\gen{su}_c(1|1)^2$   is enhanced to a Hopf algebra by introducing\footnote{With a slight abuse of notation we identify $U$ with $U\times   \id$.}
\begin{equation}\label{eq: SuperchargesCoproducts}
\Delta(Q_A)=Q_A\otimes\id+U\otimes Q_A,\qquad \Delta(S_A)=S_A\otimes\id+U^{-1}\otimes S_A,
\end{equation}
which are  the  coproducts of the supercharges   in   \emph{Drinfeld first realization}.
Acting with $\mathcal P$ provides the opposite coproducts 
\begin{equation}\label{eq: SuperchargesOpCoproducts}
\Delta^{\rm op}(Q_A)=Q_A\otimes U+\id\otimes Q_A,\qquad \Delta^{\rm op}(S_A)=S_A\otimes U^{-1}+\id\otimes S_A.
\end{equation}
The coproduct is an algebra homomorphism, therefore $\Delta(H_A),\Delta(P), \Delta(K)$ are   obtained by   anticommuting (\ref{eq: SuperchargesCoproducts}): for example, 
\begin{equation}
\Delta(H_A)=H_A\otimes\id+\id\otimes H_A,  \qquad \Delta(P)=P\otimes\id+U^2\otimes P.
\end{equation}
Using (\ref{eq: LeftSingleParticleCentralRep}) it is straightforward to check that the central charges are co-commutative.

The coproducts provide the two-particle representation of $\gen{su}_c(1|1)^2$, which is sufficient to solve the scattering problem as the system is assumed integrable and $n$-body processes factorise into 2-body ones, as allowed by the  QYBE.

%%%%%%%%%%%%%%%%%%%%%%%

\subsection{$\ads_3\times S^3\times S^3\times S^1$ R-matrix}

Using the basis $\p{\phi_L,\phi_R,\psi_L,\psi_R}$, the R-matrix for the worldsheet scattering of strings on  $\ads_3\times S^3\times S^3\times S^1$ with mixed flux assumes the form
\begin{align}
R\ket{\phi_{L\,1}\,\phi_{L\,2}}&=\alpha_{LL}\ket{\phi_{L\,1}\,\phi_{L\,2}},\qquad  R\ket{\phi_{L\,1}\,\psi_{L\,2}}=\beta_{LL}\ket{\phi_{L\,1}\,\psi_{L\,2}}+\gamma_{LL}\ket{\psi_{L\,1}\,\phi_{L\,2}},\nonumber\\
 R\ket{\psi_{L\,1}\,\phi_{L\,2}}&=\mu_{LL}\,\ket{\psi_{L\,1}\phi_{L\,2}}+\nu_{LL}\ket{\phi_{L\,1}\,\psi_{L\,2}}, \qquad R\ket{\psi_{L\,1}\,\psi_{L\,2}}=\rho_{LL}\ket{\psi_{L\,1}\,\psi_{L\,2}},
\end{align}
in the LL sector and 
\begin{align}
R\ket{\phi_{L\,1}\,\phi_{R\,2}}&=\alpha_{LR}\ket{\phi_{L\,1}\,\phi_{R\,2}}+\beta_{LR}\ket{\psi_{L\,1}\,\psi_{R\,2}},\qquad R\ket{\phi_{L\,1}\,\psi_{R\,2}}=\gamma_{LR}\ket{\phi_{L\,1}\,\psi_{R\,2}}, \nonumber\\
 R\ket{\psi_{L\,1}\,\phi_{R\,2}}&=\mu_{LR}\ket{\psi_{L\,1}\,\phi_{R\,2}}, \qquad R\ket{\psi_{L\,1}\,\psi_{R\,2}}=\nu_{LR}\ket{\psi_{L\,1}\,\psi_{R\,2}}+\rho_{LR}\ket{\phi_{L\,1}\,\phi_{R\,2}},
\end{align}
in the LR sector. According to LR symmetry, the RL and RR sectors are found by just swapping L and R in the above expressions. Then,  (\ref{eq: QuasiCocommutativity}) fixes the LL and LR scattering elements:
\begin{align}\label{eq: RmatrixEntries}
\alpha_{LL}&=\sigma_{LL} ,\qquad  \beta_{LL}=\f{\sigma_{LL}}{U_1}\,\f{x^+_{L\,2}-x^+_{L\,1}}{x^+_{L\,2}-x^-_{L\,1}},\qquad \gamma_{LL}=\sigma_{LL} \,\f{\eta_{L\,2}\p{x^+_{L\,1}-x^-_{L\,1}}}{\eta_{L\,1}\p{x^+_{L\,2}-x^-_{L\,1}}},\nonumber\\
\mu_{LL}&= \sigma_{LL}\,\f{U_2\p{x^-_{L\,2}-x^-_{L\,1}}}{x^+_{L\,2}-x^-_{L\,1}},\qquad  \nu_{LL}=\gamma_{LL},\qquad \rho_{LL}=\sigma_{LL} \,\f{U_2\p{x^+_{L\,1}-x^-_{L\,2}}}{U_1\p{x^+_{L\,2}-x^-_{L\,1}}},\nonumber\\
\alpha_{LR}&=\sigma_{LR},\qquad  \beta_{LR}=\sigma_{LR}\,\f{ \eta_{L\,1}\p{x^+_{R\,2}-x^-_{R\,2}}}{ \eta_{R\,2}\p{x^+_{L\,1}\,x^-_{R\,2}-1}},\qquad \gamma_{LR}=\sigma_{LR}\,\f{ U_1\p{x^-_{L\,1}\,x^-_{R\,2}-1}}{  x^+_{L\,1}\,x^-_{R\,2}-1},\nonumber\\
\mu_{LR}&=\f{\sigma_{LR}}{U_2}\,\f{ x^+_{L\,1}\,x^+_{R\,2}-1}{  x^+_{L\,1}\,x^-_{R\,2}-1},\qquad  \nu_{LR}=\sigma_{LR}\,\f{ U_1\p{x^-_{L\,1}\,x^+_{R\,2}-1}}{ U_2\p{ x^+_{L\,1}\,x^-_{R\,2}-1}},\qquad \rho_{LR}= \beta_{LR},
\end{align}
where $x^\pm_{A\,j}$ are the kinematic variables referring to the state $\ket{\Phi_{A\,j}}$, with $\Phi=\phi,\psi$; $A=L,R$ and $j=1,2$. For simplicity, we will consider   scattering particles with equal masses $m$, leaving $m$ unspecified. Accordingly,  $\alpha_{RL},\dots, \rho_{RR}$ are obtained by exchanging  L and R in (\ref{eq: RmatrixEntries}). As a result, the R-matrix is fixed up to   four scalar factors $\sigma_{LL},\sigma_{LR},\sigma_{RL},\sigma_{RR}$: these are the {dressing phases} determined by the crossing equations  (\ref{eq: CrossingEquations}). Such scalars encode important physical informations (\eg the bound states of the system); however, they do not affect the   Hopf algebra structure  and specifying their exact form will be unnecessary.

%%%%%%%%%%%%%%%%%%%%%%% 
%%%%%%%%%%%%%%%%%%%%%%%

\section{RTT Realization of the Yangian}\label{sec: RTTReal}

\subsection{Representations of the RTT Generators}

In the form (\ref{eq: Rmatrixglbasis}), the R-matrix entries (\ref{eq: RmatrixEntries}) are distributed as 
\begin{equation}
\begin{aligned}
R^{11}_{11}&=\alpha^{LL} , &\qquad  R^{12}_{12}&=\alpha^{LR},&\qquad  R^{13}_{13}&=\beta^{LL},&\qquad  R^{14}_{14}&=\gamma^{LR},\nonumber\\
R^{21}_{21}&=\alpha^{RL} , &\qquad  R^{22}_{22}&=\alpha^{RR},&\qquad  R^{23}_{23}&=\gamma^{RL},&\qquad  R^{24}_{24}&=\beta^{RR},\nonumber\\
R^{31}_{31}&=\mu^{LL} , &\qquad  R^{32}_{32}&=\mu^{LR},&\qquad  R^{33}_{33}&= \rho^{LL},&\qquad  R^{34}_{34}&=\nu^{LR},\nonumber\\
R^{41}_{41}&=\mu^{RL} , &\qquad  R^{42}_{42}&=\mu^{RR},&\qquad  R^{43}_{43}&=\nu^{RL},&\qquad  R^{44}_{44}&=\rho^{RR},\nonumber\\
R^{12}_{34}&=-\beta^{LR} , &\qquad  R^{13}_{31}&=-\gamma^{LL},&\qquad  R^{21}_{43}&=-\beta^{RL},&\qquad  R^{24}_{42}&=-\gamma^{RR},\nonumber\\
R^{31}_{13}&=\nu^{LL} , &\qquad  R^{34}_{12}&=\rho^{LR},&\qquad  R^{42}_{24}&=\nu^{RR},&\qquad  R^{43}_{21}&=\rho^{RL}.
\end{aligned}
\end{equation}
In a neighbourhood of $u_j=\infty$, the $x^\pm_{A\,j}$ given by 
\begin{equation}\label{eq: ZhukovskyAsympt}
x^\pm_{A\,j}=u_j +\f{\pm\im m +2\kappa_A}h+\f{-h^2\pm{2\im m\,\kappa_A} }{h^2\,u_j}+\f{\pm\im m +2\kappa_A}{h\,u_j^2}+\dots,\qquad A=L,R
\end{equation}
satisfy (\ref{eq: ZhukovskyConstr}). The spectral parameters $u_j$ expressed in terms of $x^\pm_{A\,j}$ are then
\begin{equation}
\f12\p{x^+_{A\,j}+\f1{x^+_{A\,j}}+x^-_{A\,j}+\f1{x^-_{A\,j}}}=u_j+\f{2\,\kappa_A}{h},\qquad A=L,R.
\end{equation}
Thanks to (\ref{eq: ZhukovskyAsympt}) one can expand the R-matrix and read off the algebra commutation relations from (\ref{eq: RTTRelationsComponents}) and the representations of their generators  from  (\ref{eq: RTTRepsfromRmatrix}). To begin with, we find
\begin{equation}
T^{\,a}_{(-1)\,b}={\delta^a}_b\,U_2^{|b|},  
\end{equation}
namely $T^{\,1}_{(-1)\,1}=\id$, $T^{\,1}_{(-1)\,2}=0$, $T^{\,3}_{(-1)\,3}=U_2\,\id$ and so forth. As anticipated, the leading term in the asymptotic expansion of $\mathcal T(u)$ turns out to be central. Since we are embedding $\gen{su}_c(1|1)^2$ in $\gen{gl}(2|2)$, only 8 of the 16 $T^{\,a}_{(0)\,b}$ are non-vanishing. The same happens for the  $T^{\,a}_{(n)\,b}$ with $n\geq0$: these are $T^{\,a}_{(n)\,a},T^{\,a+2}_{(n)\,a+2}, T^{\,a}_{(n)\,a+2}, T^{\,a+2}_{(n)\,a}$ with $a=1,2$. A sample of these are
\begin{align}
T^{\,1}_{(0)\,1}&=\f{\im m}{h}\p{{e^2}_2-{e^3}_3},\qquad T^{\,1}_{(0)\,3}=-\f{2\im\sqrt m\,\eta_{L\,2}}{h}\,{e^1}_3+\f{\sqrt m\p{1-U_2^2}}{\eta_{R\,2}}\,{e^4}_2,\nonumber\\
T^{\,2}_{(0)\,2}&=\f{\im m}{h}\p{{e^1}_1-{e^4}_4},\qquad T^{\,2}_{(0)\,4}=-\f{2\im\sqrt m\,\eta_{R\,2}}{h}\,{e^2}_4+\f{\sqrt m\p{1-U_2^2}}{\eta_{L\,2}}\,{e^3}_1.
\end{align}
The expressions of the higher generators are rather lenghty, \eg
\begin{align}
 T^{\,1}_{(+1)\,1}&= \frac{m}{2h^2\,\text{$\eta_{R\,2}$}^2} \left[-\p{m-4 \im \,\kappa }\text{$\eta_{R\,2}$}^2 +{2h^2\, U_2 \left(-1+U_2^2\right)}\right]\,{e^2}_2+\nonumber\\
 &-\frac{m
\left(m-4 \im \, \kappa -U_2 \left(8\, \text{$\eta_{L\,2} $}^2+(m-4 \im \, \kappa ) U_2\right)\right)}{2 h ^2 \left(-1+U_2^2\right)}\,{e^3}_3.
\end{align}

%%%%%%%%%%%%%%%%%%%%%%%

\subsection{Abstract Commutation Relations}

Expanding (\ref{eq: RTTrelations}) around $\infty$ with respect to both $u_1$ and $u_2$ gives the abstract commutation relations for the $\bb T$s. A sample of the latter is
\begin{align}\label{eq: exofTsCommRel}
\acomm{\bb  T^{\,4}_{(1)\,2},\bb  T^{\,3}_{(0)\,1}}&=\f{\im m}h\,\bb  T^{\,3}_{(0)\,1}\bb  T^{\,4}_{(0)\,2}+ \f{2\im m}h\p{ \bb  T^{\,2}_{(0)\,2}-\bb U\,\bb  T^{\,4}_{(0)\,4}}+\f{m^2}{h^2}\p{\f{2\im \kappa}m +1 }\p{\id-\bb U^2},\nonumber\\
 \comm{\bb  T^{\,3}_{(0)\,1},\bb  T^{\,2}_{(0)\,2}}& =\comm{\bb  T^{\,3}_{(0)\,1},\bb  T^{\,4}_{(0)\,4}}=\f{\im m}h\,\bb  T^{\,3}_{(0)\,1},\qquad  \acomm{\bb  T^{\,4}_{(0)\,2},\bb  T^{\,3}_{(0)\,1}}=\f{2\im m}h\p{\id-\bb U^2}.
\end{align}
Notice that  $m,h$ and $\kappa$ combine to   structure constants. The commutation rules of  $\bb T$s  provide   $\bb J$s': for instance, the relations (\ref{eq: exofTsCommRel}) imply
\begin{align}
\acomm{\bb  J^{\,4}_{(1)\,2},\bb  J^{\,3}_{(0)\,1}}=-\f{\im \kappa}{2m}\p{\id-\bb U^2}+\f12\p{\id +\bb U^2}\p{\bb  J^{\,2}_{(0)\,2}-\bb  J^{\,4}_{(0)\,4}}.
\end{align}

%%%%%%%%%%%%%%%%%%%%%%%

\subsection{Lie Superalgebra from RTT}

We now reconstruct the whole Hopf algebra $\mathcal A$. By choosing 
\begin{align}\label{eq: LfromJ}
 \bb Q_L&= \sqrt m \,\bb J^{\,3}_{(0)\,1},\qquad  \bb S_L= \sqrt m\, { \bb J^{\,1}_{(0)\,3}},\qquad  \bb H_L= m\p{\bb  J^{\,1}_{(0)\,1}- \bb J^{\,3}_{(0)\,3}},\qquad  \bb P=\f{\im h}{2}\p{ \bb U^2-1},\nonumber\\
 \bb Q_R&=\sqrt m \, \bb J^{\,4}_{(0)\,2},\qquad  \bb S_R= \sqrt m \, \bb J^{\,2}_{(0)\,4},\qquad  \bb H_R=m\p{ \bb  J^{\,2}_{(0)\,2}- \bb J^{\,4}_{(0)\,4}},\qquad  \bb K=\f{\im h}{2}\p{1- \bb U^{-2}},
\end{align}
we exactly reproduce the $\gen{su}_c(1|1)^2$ graded commutation relations:
\begin{equation}
\acomm{ \bb Q_A, \bb S_B}=\delta_{AB}\,  \bb H_B,\qquad \acomm{ \bb Q_L, \bb Q_R}=   \bb P,\qquad \acomm{ \bb S_L, \bb S_R}=   \bb K;\qquad A,B=L,R.
\end{equation}

The   coproducts (\ref{eq: SuperchargesCoproducts}) are recovered by applying $\Delta$ to (\ref{eq: LfromJ}) by means of  (\ref{eq: Jcoproduct}). As for the antipodes, the equations   (\ref{eq: Jantipode}) provide
\begin{align}
\Sigma\comm{\bb Q_A}&=-\bb U^{-1}\,\bb Q_A,\qquad \Sigma\comm{\bb S_A}=-\bb U\,\bb S_A,\qquad  \Sigma\comm{\bb H_A}=- \bb H_A,\nonumber\\
  \Sigma\comm{\bb P}&=-\bb U^{-2}\, \bb P,\qquad \Sigma\comm{\bb K}=-\bb U^{2}\, \bb K,\qquad A=L,R,
\end{align}
which   preserve the algebra commutation relations. Notice that the antipode is involutive when evaluated on $\gen{su}_c(1|1)^2$.

%%%%%%%%%%%%%%%%%%%%%%%

\subsection{Outer Generators}

As in \cite{Beisert:2014hya,Hoare:2015kla}, the sum
\begin{equation}
\bb A= \bb  J^{\,1}_{(0)\,1}+\bb  J^{\,2}_{(0)\,2}+ \bb J^{\,3}_{(0)\,3}+ \bb J^{\,4}_{(0)\,4}
\end{equation}
is an outer central generator of $\gen{su}_c(1|1)^2$. Since $\bb A$ never appears on either the left or the right hand side of the non-trivial $\mathcal A$'s commutation relations, it can be modded out of the algebra. In other words, if $\bb B_L$ and $\bb B_R$ are the bosonic generators
\begin{equation}
\bb B_L=  \bb  J^{\,1}_{(0)\,1}+ \bb J^{\,3}_{(0)\,3} ,\qquad  \bb B_R=  \bb  J^{\,2}_{(0)\,2}+ \bb J^{\,4}_{(0)\,4},\qquad \bb B_L + \bb B_R =\bb A,
\end{equation}
whose action on the supercharges reads\footnote{LR symmetry provides the other commutation relations, \eg $\comm{\bb B_R,\bb Q_L}$ and  $\comm{\bb B_R,\bb Q_R}$.}
\begin{equation}
\comm{\bb B_L,\bb Q_L}=  \bb Q_L,  \qquad  \comm{\bb B_L,\bb Q_R}=  -\bb Q_R,\qquad  \comm{\bb B_L,\bb S_L}=  -\bb S_L,  \qquad  \comm{\bb B_L,\bb S_R}=  \bb S_R,
\end{equation}
one realises that $ \comm{\bb B_L,\cdot}$ and $\comm{\bb B_R,\cdot}$ are not   independent: in fact, they add up to $ \comm{\bb A,\cdot}\equiv 0$. Then, the  Lie symmetry algebra of the theory is not the full $\gen{gl}(1|1)^2$ but, rather, $\gen{gl}(1|1)^2/\bb A$.  

%%%%%%%%%%%%%%%%%%%%%%%

\subsection{$\mathcal Y[\gen{su}_c(1|1)^2]$ Yangian from RTT}

Setting 
\begin{align}
\hat{ \bb Q}_L&=\sqrt m \, \bb J^{\,3}_{(+1)\,1}+\f\kappa h\, { \bb Q}_L+\f12 \p{1+ \bb U^2} { \bb S}_R,\qquad \hat{ \bb S}_L= \sqrt m \,\bb J^{\,1}_{(+1)\,3}+\f\kappa h\,{ \bb S}_L+\f12 \p{1+ \bb U^{-2}}  { \bb Q}_R,\nonumber\\
\hat{ \bb H}_L&=m\p{ \bb J^{\,1}_{(+1)\,1}- \bb J^{\,3}_{(+1)\,3}}+\f{2\kappa} h\, \bb H_L+\f{\im h}{4  }\p{ \bb U^{2}- \bb U^{-2}},\nonumber\\
\hat{ \bb Q}_R&= \sqrt m \,\bb J^{\,4}_{(+1)\,2}-\f\kappa h\,  { \bb Q}_R+\f12 \p{1+ \bb U^2} { \bb S}_L,\qquad\hat{ \bb S}_R= \sqrt m \,\bb J^{\,2}_{(+1)\,4}-\f\kappa h\, { \bb S}_R+\f12 \p{1+ \bb U^{-2}} { \bb Q}_L,\nonumber\\
\hat{ \bb H}_R&=m\p{ \bb J^{\,2}_{(+1)\,2}- \bb J^{\,4}_{(+1)\,4}}-\f{2\kappa} h\, \bb H_R+\f{\im h}{4  }\p{ \bb U^{2}- \bb U^{-2}},
\end{align}
and
\begin{equation}
\hat{ \bb P}=\f12\p{1+ \bb U^2}\p{ \bb H_L+ \bb H_R},\qquad \hat{ \bb K}=\f12\p{1+ \bb U^{-2}}\p{ \bb H_L+ \bb H_R}
\end{equation}
we obtain the commutation relations for the level one of the $\mathcal Y[\gen{su}_c(1|1)^2]$ Yangian:
\begin{align}
\acomm{\hat{ \bb Q}_A, \bb S_B}&=\acomm{{ \bb Q}_A,\hat{ \bb S}_B}=\delta_{AB}\,\hat{ \bb H}_B,\qquad A,B=L,R,\nonumber\\
\acomm{\hat{ \bb Q}_L, \bb Q_R}&=\acomm{{ \bb Q}_L,\hat{ \bb Q}_R}=\hat{ \bb P},\qquad \acomm{\hat{ \bb S}_L, \bb S_R}=\acomm{{ \bb S}_L,\hat{ \bb S}_R}=\hat{ \bb K}.
\end{align}
The corresponding coproducts are
\begin{align}
\Delta(\hat{ \bb Q}_L)&=\hat{ \bb Q}_L\otimes\id+\bb U\otimes \hat{ \bb Q}_L+\f{\im  }h( { \bb Q}_L\otimes\bb H_L-\bb U\,\bb H_L\otimes { \bb Q}_L+\nonumber\\
&+\f{\bb P}{\bb U}\otimes { \bb S}_R-\bb U^2\, { \bb S}_R\otimes \bb P),\nonumber\\
\Delta(\hat{ \bb S}_L)&=\hat{ \bb S}_L\otimes\id+\bb U^{-1}\otimes \hat{ \bb S}_L+\f{\im }h( {\bb U^{-1}}\,{\bb H_L}\otimes { \bb S}_L-{ \bb S}_L\otimes\bb H_L+\nonumber\\
&+\f{{ \bb Q}_R}{\bb U^2}\otimes \bb K-\bb U\, \bb K\otimes { \bb Q}_R),\nonumber\\
\Delta(\hat{ \bb H}_L)&=\hat{ \bb H}_L\otimes\id + \id \otimes \hat{ \bb H}_L+\f{\im }h\p{ \bb U^{-2}\,{ \bb P}\otimes { \bb K}-\bb U^{2}\,{ \bb K}\otimes { \bb P}},
\end{align}
 and $\Delta\p{\gen J_R}=\Delta\p{\gen J_L}_{L\to R}$ for the R generators. Finally, $\Delta$ acts on the Yangian central charges as
\begin{align}
\Delta(\hat{ \bb P} )&=\hat{ \bb P} \otimes\id + \bb U^2 \otimes \hat{ \bb P} +\f{\im  }h\p{   \bb P \otimes  \bb H-\bb U^2\, \bb H\otimes  \bb P},\nonumber\\
\Delta(\hat{ \bb K} )&=\hat{ \bb K} \otimes\id + \bb U^{-2} \otimes \hat{ \bb K} +\f{\im  }h\p{   \bb U^{-2}\,\bb H \otimes  \bb K-  \bb K\otimes  \bb H}.
\end{align}
Ultimately, the antipodes are
\begin{align}
\Sigma[{\hat{\bb Q}_A}]&=-\bb U^{-1}\,\hat{\bb Q}_A,\qquad \Sigma[{\hat{\bb S}_A}]=-\bb U\,\hat{\bb S}_A,\qquad  \Sigma[{\hat{\bb H}_A}]=- \hat{\bb H}_A,\nonumber\\
  \Sigma[{\hat{\bb P}}]&=-\bb U^{-2}\, \hat{\bb P},\qquad \Sigma[{\hat{\bb K}}]=-\bb U^{2}\, \hat{\bb K},\qquad A=L,R.
\end{align}
The antipode $\Sigma$ acts as an involution on the level one partners of the $\gen{su}_c(1|1)^2$ generators. This will no longer be true for the secret symmetry, as we shall see in the next subsection.

%%%%%%%%%%%%%%%%%%%%%%%

\subsection{Secret Symmetry}

The outer generator
\begin{equation}
  \ss=  {\bb  J}^{\,1}_{(+1)\,1}+ {\bb  J}^{\,2}_{(+1)\,2}+ { \bb J}^{\,3}_{(+1)\,3}+  {\bb J}^{\,4}_{(+1)\,4}
\end{equation}
is not central and cannot be quotiented out as it was previously done for $\bb A$. As a result, the bosonic charges 
\begin{equation}
\ss_L=  {\bb  J}^{\,1}_{(+1)\,1} + { \bb J}^{\,3}_{(+1)\,3} ,\qquad \ss_R=   {\bb  J}^{\,2}_{(+1)\,2} +  {\bb J}^{\,4}_{(+1)\,4}
\end{equation}
are   independent. Their commutation relations read
\begin{align}
\comm{\ss_L, \bb Q_L}&=2\,\hat{\bb Q}_L-\p{\id+\bb U^2}\bb S_R,\qquad \comm{\ss_L, \bb Q_R}= -\p{\id+\bb U^2}\bb S_L,\nonumber\\
\comm{\ss_L, \bb S_L}&=-2\,\hat{\bb S}_L+\p{\id+\bb U^{-2}}\bb Q_R,\qquad \comm{\ss_L, \bb S_R}= \p{\id+\bb U^{-2}}\bb Q_L, 
\end{align}
where the others are given by the swapping L $\leftrightarrow$ R. The  coproducts are
\begin{equation}
\Delta(\ss_L)=\ss_L\otimes \id + \id\otimes \ss_L+\f{2\im  }h\p{\bb U\,\bb S_L\otimes \bb Q_L+\bb U^{-1}\bb Q_L\otimes \bb S_L}
\end{equation}
and $\Delta(\ss_R)=\Delta(\ss_L)_{L\to R}$, while the antipodes read
\begin{equation}
\Sigma[\ss_L]=-\ss_L+(2\im  /h)\,\bb H_L,\qquad \Sigma[\ss_R]=\Sigma[\ss_L]_{L\to R}.
\end{equation}
The existence of these additional generators implies that the symmetry of the theory is not just $\mathcal Y[\gen{su}_c(1|1)^2]$, but rather $\mathcal A=\mathcal Y[\gen{gl}_c(1|1)^2/\bb A]$. These $\ss_L,\ss_R$ generate the $\ads_3\times S^3\times S^3\times S^1$ version of the \emph{secret symmetry} first found in the $\ads_5\times S^5$ case \cite{Matsumoto:2007rh, Regelskis:2011fa, deLeeuw:2011fr } and then discovered also in the worldsheet scattering on $\ads_3\times S^3\times \mathcal M^4$ with  RR flux \cite{Pittelli:2014ria} as well as in the massive sector of the $\ads_2\times S^2$ superstring \cite{Hoare:2015kla}. On the field theory side, the secret symmetry corresponds to an helicity operator acting on scattering amplitudes \cite{Beisert:2011pn} as well as on Wilson loops \cite{Munkler:2015xqa, Klose:2016qfv,Dekel:2016oot}. On the string theory side, the existence of  secret symmetries  was demonstrated by means of the pure spinor formalism  \cite{Berkovits:2011kn}. As anticipated, the antipode is not involutive  on  $\ss_A$:
\be
\Sigma^2(\ss_A)=\ss_A-(4\im/h)\,\bb H_A,\qquad A=L,R.
\ee
This signals that $\mathcal A$  possesses a non-trivial \emph{Liouville contraction} $\mathcal Z(u)$ \cite{Chari:1994pz, Kassel:1995xr, Beisert:2014hya} shifting  the double antipode of diagonal elements, $\eg$ ${\bb  J}^{\,a}_{(+1)\,a}, {\bb  J}^{\,a}_{(+2)\,a}$. More precisely, $\mathcal Z(u)$ is a central element of the algebra satisfying $\id\otimes \mathcal Z(u)=\mathcal T(u)\Sigma\comm{\mathcal T(u)}$. In our case, by direct computation we find 
\begin{equation}
  \mathcal Z_A(u)=\exp\acomm{(2\im/h)\,u^{-2}\,\bb H_A+\dots}.
  %\mathcal Z_A(u)=\id -(4 \,m / h^2)\,u^{-2}  \,\bb H_A+\dots,\qquad A=L,R.
\end{equation}
The   order $u^{-2}$ deviation from $ \mathcal Z_A(u)=\id$ is proportional to the shift in $\Sigma^2(\ss_A)$, while higher orders provide $\Sigma^2({\bb  J}^{\,a}_{(+2)\,a})$, $ \Sigma^2({\bb  J}^{\,a}_{(+3)\,a})$ and so forth. 

 %%%%%%%%%%%%%%%%%%%%%%%
 
 \subsection{Evaluation Representation}
 
 Acting with the  function $\rho_u$ on the Yangian charges   one finds that the latter  are in evaluation representation. For instance, if $  Q_L=\rho_u\,   {\bb Q}_L$ and $\hat Q_L=\rho_u\,\hat  {\bb Q}_L$, then $\hat Q_L \ket{\psi_R} = \comm{u - (2\,\kappa/h)}Q_L \ket{\psi_R}$ holds. In general, if $\hat{\gen J}$ is the level one counterpart of $ {\gen J}$, one obtains
\be 
 \hat{\gen J} \ket{\Phi_A} = \p{u + \f{2\,\kappa_A}{h}}\gen J \ket{\Phi_A},\qquad \Phi=\phi,\psi,\qquad  A=L,R.
\ee
Then, the quantised coefficient of the Wess-Zumino term $\kappa$ splits the evaluation representation in two branches: the left one, with effective spectral parameter $u_L=u + (2\,\kappa/h)$; and the right one, with $u_R=u - (2\,\kappa/h)$. In the limit $\kappa\to0$, corresponding to   RR flux only, the two branches collapse and become one, with a single spectral parameter $u$.  This was the situation analysed in \cite{Pittelli:2014ria, Regelskis:2015xxa}.

%%%%%%%%%%%%%%%%%%%%%%%
%%%%%%%%%%%%%%%%%%%%%%%

\section{Loop Algebra and Universal Classical r-matrix}\label{sec: Classicalrmatrix}

\subsection{Loop Algebra}

In this section we study the  \emph{classical limit} of the $\ads_3\times S^3\times S^3\times S^1$ R-matrix and its Yangian   by taking the effective string tension $h$ to be very large, namely\footnote{By AdS/CFT, this limit corresponds  to the strong coupling limit on the gauge theory side.} $ h\to\infty$. To this aim, we parametrise the Zhukovsky variables $x^\pm_{L,R}$  in terms of the constants $h,m$ and of a new variable $z$ \cite{Arutyunov:2006iu}:
\begin{equation}\label{eq: zhukclassexp}
x^\pm_L=x^\pm_R=z\p{\sqrt{1-\f{( m/h)^2}{(z-z^{-1})^2}}\pm \frac{\im ( m/h)}{z-z^{-1}}}.
\end{equation}
Consequently, we expand with respect to $ h^{-1}$. Such $x^\pm_{L,R}$ satisfy the constraint (\ref{eq: ZhukovskyConstr}) because $\kappa$-corrections are of order $ h^{-2}$. In this regime, the spectral parameter $u$ reads $u=z+z^{-1}$ and the representations of the Yangian generators become
\begin{align}\label{eq: classicalgens}
\f{\gen Q_L }{\sqrt m}& = - \f{{e^2}_4+z\, {e^3}_1}{\sqrt{z^2-1}},\qquad  \f{\gen S_L}{\sqrt m} = - \f{z\,{e^1}_3+\, {e^4}_2}{\sqrt{z^2-1}},\qquad \f{\gen H_L}m=\f{z^2\p{{e^1}_1-{e^3}_3}+\p{{e^2}_2-{e^4}_4}}{1-z^2},\nonumber\\
\f{\gen Q_R }{\sqrt m} & = - \f{{e^1}_3+z\, {e^4}_2}{\sqrt{z^2-1}},\qquad  \f{\gen S_R }{\sqrt m} = - \f{z\,{e^2}_4+ {e^3}_1}{\sqrt{z^2-1}},\qquad \f{\gen H_R }{  m}=\f{\p{{e^1}_1-{e^3}_3}+z^2\p{{e^2}_2-{e^4}_4}}{1-z^2},\nonumber\\
 \gen B_L&=u^{-1}\,\ss_L=\p{z^4-1}^{-1}\comm{z^4\,{e^1}_1+{e^2}_2-z^2\p{3z^2-2}{e^3}_3 + \p{2z^2-3}{e^4}_4  } ,\nonumber\\
\gen  B_R &=u^{-1}\,\ss_R=\p{z^4-1}^{-1}\comm{ {e^1}_1+z^4\,{e^2}_2+\p{2z^2-3}{e^3}_3 -z^2 \p{3z^2-2}{e^4}_4  }.
\end{align}

Let $\gen G$ be one of the generators reported above.  Then,  we can uplift $\gen G$ to its level $n$ counterpart   via evaluation representation: $\gen G^{(n)}=u^n\,\gen G$. The  algebra formed by (\ref{eq: classicalgens}) is then
\begin{align}\label{eq: gl112loopalgebra}
&\acomm{\gen Q_L^{( n_1)},\gen S_L^{( n_2)}}=\gen H_L^{( n_1+n_2)},\qquad \acomm{\gen Q_L^{( n_1)},\gen Q_R^{( n_2)}}=\acomm{\gen S_L^{( n_1)},\gen S_R^{( n_2)}}=\gen H_L^{( n_1+n_2-1)}+\gen H_R^{( n_1+n_2-1)},\nonumber\\
&\comm{\gen B_L^{( n_1)},\gen Q_L^{( n_2)}}=2\,\gen Q_L^{(n_1+n_2)}-2\,\gen S_R^{( n_1+n_2-1)},\qquad \comm{\gen B_L^{( n_1)},\gen S_L^{( n_2)}}=-2\,\gen S_L^{(n_1+n_2)}+2\,\gen Q_R^{( n_1+n_2-1)},\nonumber\\
&\comm{\gen B_L^{( n_1)},\gen Q_R^{( n_2)}}=-2\,\gen S_L^{(n_1+n_2-1)} ,\qquad \comm{\gen B_L^{( n_1)},\gen S_R^{( n_2)}}=2\,\gen Q_L^{(n_1+n_2-1)} ,\qquad n_1,n_2\in\bb Z.
\end{align}
  The remaining  commutation rules descend from LR symmetry. The relations reported in (\ref{eq: gl112loopalgebra}) can be identified as a deformation of the $\gen{u}(1|1)[u,u^{-1}] \oplus \gen{u}(1|1)[u,u^{-1}] $ loop algebra {  \cite{Borsato:2017icj}}.  Let us now take the classical limit of the R-matrix and check that not only it enjoys the symmetry algebra (\ref{eq: gl112loopalgebra}), but  it can also be written as a tensor product of     $\gen{u}(1|1)[u,u^{-1}]^2$ generators.

%%%%%%%%%%%%%%%%%%%%%%%

\subsection{Classical r-matrix}

Using (\ref{eq: zhukclassexp}), the R-matrix expands as
\begin{equation}
R=\id\otimes\id +  h^{-1} r+\mathcal O( h^{-2}),
\end{equation}
where $r$ is the \emph{classical r-matrix} of the system satisfying the \emph{classical} Yang-Baxter equation
\begin{equation}\label{eq: cybe}
\comm{r_{12},r_{13}}+\comm{r_{12},r_{23}}+\comm{r_{13},r_{23}}=0,
\end{equation}
representing the QYBE (\ref{eq: qybe}) in the $h\to\infty$ regime. Since
\begin{equation} 
\Delta (\gen J)=\Delta_0 (\gen J)+h^{-1}\Delta_1 (\gen J)+\dots,\qquad \Delta^{\rm op} (\gen J)=\Delta_0 (\gen J)+h^{-1}\Delta^{\rm op}_1 (\gen J)+\dots,
\end{equation}  
quasi co-commutativity (\ref{eq: QuasiCocommutativity}) becomes
\begin{equation}\label{eq: ClassicalQuasiCocommutativity}
\comm{\Delta_0 (\gen J),r}=\Delta_1 (\gen J)-\Delta^{\rm op}_1 (\gen J),\qquad \forall\,\gen J\in \mathcal A.
\end{equation} 
Therefore, the right-hand side of (\ref{eq: ClassicalQuasiCocommutativity}), also known as the \emph{cobracket} of $\gen J$, has to match the commutator $\comm{\Delta_0 (\gen J),r}$. The classical r-matrix obtained from   (\ref{eq: RmatrixEntries}) reads
\begin{align}\label{eq: evrepclassrmatrix}
r&=\frac{2{\rm i}}{u_1-u_2}[ \gen Q_L\otimes \gen S_L-\gen S_L\otimes \gen Q_L+\gen Q_R\otimes \gen S_R-\gen S_R\otimes \gen Q_R+\nonumber\\
&-\frac1{4u_1} (\ss_L-\ss_R)\otimes (\gen H_L - \gen H_R)-\frac1{4u_2}(\gen H_L-\gen H_R)\otimes(\ss_L-\ss_R)+\nonumber\\
&-\frac1{4u_2} (\ss_L+\ss_R)\otimes (\gen H_L + \gen H_R)-\frac1{4u_1}(\gen H_L+\gen H_R)\otimes(\ss_L+\ss_R)].
\end{align}
The generators $\gen Q_L,\gen S_L,\dots, \ss_L,\ss_R$ are genuine symmetries of the r-matrix: for instance,
\begin{align}
\comm{\Delta_0(\ss_L),r}=4\im \p{\gen Q_L\otimes \gen S_L+\gen S_L\otimes \gen Q_L}=\Delta_1(\ss_L)-\Delta_1^{\rm op}(\ss_L),
\end{align}
which agrees with (\ref{eq: ClassicalQuasiCocommutativity}). In the region $|u_2|<|u_1|$, the   r-matrix can  be rewritten as\footnote{The $r$-matrix for  $|u_1|<|u_2|$ is obtained by making the  replacement   $\gen G_1^{(-1-n)}\otimes \gen G_2^{(n)}\to -\, \gen G_1^{( n)}\otimes \gen G_2^{(-1-n)}$ in (\ref{eq: crmrepindep}).} 
\begin{align}\label{eq: crmrepindep}
r&=  2\im\,\left[\gen Q_L^{(-1-n)}\otimes \gen S_L^{(n)}-\gen S_L^{(-1-n)}\otimes \gen Q_L^{(n)}+\gen Q_R^{(-1-n)}\otimes \gen S_R^{(n)}-\gen S_R^{(-1-n)}\otimes \gen Q_R^{(n)}\right.+\nonumber\\
&\left.- \tilde{\gen L}^{(-1-n)}\otimes \gen M^{(n)}-\gen M^{(-1-n)}\otimes\tilde{\gen L}^{(n)}-  \tilde{\gen B}^{( -n)}\otimes \gen H^{(n-1)}-\gen H^{(-2-n)}\otimes\tilde{\gen B}^{(n+1)}\right]  ,
\end{align}
where     the redefinitions 
\begin{align}
\gen H^{(0)}&=(\gen H_L + \gen H_R)/2,\qquad \gen B^{(1)}=(\ss_L+\ss_R)/2,\qquad \tilde{\gen B}^{(1)}=\gen B^{(1)}+\gen H^{(0)}-2\gen H^{(-1)},\nonumber\\
\gen M^{(0)}&=(\gen H_L - \gen H_R)/2,\qquad \gen L^{(1)}=(\ss_L-\ss_R)/2,\qquad \tilde{\gen L}^{(1)}=\gen L^{(1)}+\gen M^{(0)},
\end{align}
were used. The object in (\ref{eq: crmrepindep}) is representation independent; therefore, it is a well-posed  candidate for the universal r-matrix of the loop algebra (\ref{eq: gl112loopalgebra}). In the light of (\ref{eq: crmrepindep}), (\ref{eq: evrepclassrmatrix}) is in fact in evaluation representation.

 Finally, notice that  the identifications 
\begin{align}
 \gen Q_A^{(n)}&=- {\bf Q}_{A,n},\qquad  \gen S_A^{(n)}= \bar{\bf Q}_{A,n},\qquad 2\, \gen H^{(n)}=- {{\bf H}_n },\nonumber\\
  2\, \gen M^{(n)}&=- {  {\bf M}_n},\qquad  \gen B^{(n)}=- {{\bf B}_n },\qquad 2\, \gen L^{(n)}=- {  {\bf b}_n},
\end{align}
map the r-matrix (\ref{eq: crmrepindep}) to the one found in \cite{Borsato:2017icj} up to central shifts.

%%%%%%%%%%%%%%%%%%%%%%%
%%%%%%%%%%%%%%%%%%%%%%%

\section{Conclusions}

\subsection{Discussion}

In this work we have investigated the integrability of type IIB superstrings on $\ads_3\times S^3\times S^3\times S^1$ background with mixed RR and NSNS flux from an algebraic viewpoint. Our analysis has shown  that  the worldsheet scattering  of the theory enjoys an infinite dimensional symmetry $\A$ endowed with a Hopf algebra structure and  spanned by an endless tower of generators.  Specifically, we have used the R-matrix to write down the RTT relations and  expanded these with respect to the spectral parameter, obtaining   the abstract commutation relations of $\A$. Furthermore, expanding the R-matrix itself  has provided the representations of the $\A$ generators.  We have observed that the presence of both RR and NSNS fluxes deforms not only the algebra representations but also the abstract commutation relations of the RTT generators.

The infinite dimensional algebra $\A$ is organised in levels: the bottom level, $\A_{\,-1}$,  only contains the identity  and a braiding factor $\bb U$. The latter is a   central element  deforming   commutation relations,   coproducts,   antipodes and so forth. The next level, $\A_{0}$,   coincides with the building block of the light-cone off-shell symmetry algebra of the model,  $\gen{su}_c(1|1)^2$, whose central charges depend on the braiding factor $\bb U$  and vanish in the limit $\bb U\to\id$. The level zero of $\A$ also includes  two outer generators of $\gen{su}_c(1|1)^2$, $\bb B_L,\bb B_R$. These add up to a central generator $\bb A$ that never appears in the commutation relations of $\A$, implying that  $\bb B_L,\bb B_R$   actually represent one single generator. We have summarised this situation by referring to  $\A_{0}$ as $\gen{gl}_c(1|1)^2/\bb A$. Subsequently, we have demonstrated that the level one of $\A$, $\A_{1}$, contains   level one charges of the Yangian  $\mathcal Y[\gen{su}_c(1|1)^2]$. Such charges are in evaluation representation, where each copy of $\gen{su}_c(1|1)$ presents    its own spectral parameter depending  on the quantised coefficient of the Wess-Zumino term. We have also found that at level one there are two  bonus generators, $\ss_L,\ss_R$, mapping each level of $\A$ to the next one. In contrast to $\bb B_L$ and $\bb B_R$, $\ss_L$ and $\ss_R$ are algebraically independent and  span  the $\ads_3\times S^3\times S^3\times S^1$   counterpart of the secret symmetry found in $\ads_5$ and $\ads_2$, as well as in another $\ads_3$ case.   We have   checked  that the antipode fails to be involutive when evaluated on  $\ss_L,\ss_R$, and   displayed that the failure  is quantitatively related to the non-trivial center of   $\A$.  By taking into account this secret symmetry, we have concluded  that the worldsheet scattering matrix of superstrings on $\ads_3\times S^3\times S^3\times S^1$   with mixed flux is invariant under the action of the  $\mathcal Y[\gen{gl}_c(1|1)^2/\bb A]$ Yangian, with  evaluation representation depending on the quantised coefficient of the Wess-Zumino term. 

 Finally,  taking  the large effective string tension limit, we have showed that    $\mathcal Y[\gen{gl}_c(1|1)^2/\bb A]$   reduces to a deformation of the $\gen{u}(1|1)[u,u^{-1}]^2$ loop algebra. Then, we have formulated a candidate for the universal, representation independent  $\gen{u}(1|1)[u,u^{-1}]^2$  r-matrix, linked to the  classical  limit of the R-matrix by means of the evaluation representation. We have explicitly verified that such a universal  r-matrix, in analogy to those found in \cite{Moriyama:2007jt, Beisert:2007ty,Hoare:2015kla},   is classically co-commutative with respect to the $\gen{u}(1|1)[u,u^{-1}]^2$ generators  regardless the representation.

%%%%%%%%%%%%%%%%%%%%%%%

\subsection{Outlook}

It would be very interesting to find a universal R-matrix for the worldsheet scattering on $\ads_3\times S^3\times S^3\times S^1$ with both RR and NSNS fluxes  in terms of a Drinfeld quantum double construction   \cite{Beisert:2016qei, Beisert:2017xqx}. Furthermore, it would be fascinating to see how  $q$-Poincar\'e symmetry  \cite{Stromwall:2016dyw,Borsato:2017lpf, Borsato:2017icj} behaves in presence of   mixed flux. Another intriguing   direction would be translating the Yangian symmetry derived in this paper into constraints applicable on   observables of  the conformal field theory dual \cite{Tong:2014yna,Baggio:2017kza, Eberhardt:2017fsi, Eberhardt:2017pty }, reverse-engineering what it was done in the context of $\mathcal N=4$ super Yang-Mills theory \cite{Muller:2013rta, Beisert:2015jxa, Beisert:2015uda}. Finally, it would be exciting  to extend the analysis of this paper to other supergravity backgrounds;  for instance, by extracting the Yangian corresponding to the $\ads_2\times S^2\times S^2\times T^4$ superstring \cite{Gauntlett:1997pk, Wulff:2016vqy}.

%%%%%%%%%%%%%%%%%%%%%%%

\subsection{Acknowledgements}

I would very much like to thank Riccardo Borsato, Joakim Str{\"o}mwall and Alessandro Torrielli  for valuable comments on the draft and for sharing  a copy of  \cite{Borsato:2017icj} whilst in preparation.  I am extremely  grateful to Lorenz Eberhardt,  Guido Festuccia, Olaf Lechtenfeld and Marius de Leeuw for illuminating discussions. This work was supported by the Riemann Fellowship and by the ERC STG Grant 639220.

%%%%%%%%%%%%%%%%%%%%%%%
%%%%%%%%%%%%%%%%%%%%%%%

\bibliographystyle{unsrt}
\bibliography{pitt_int}

\end{document}